\documentclass[preprint,aps,showpacs,prc,nofootinbib]{revtex4-1}
\usepackage{epsfig}

\newcommand{\BV}{\left(\begin{array}{c}}
\newcommand{\EV}{\end{array}\right)}
\newcommand{\BM}{\left(\begin{array}{cc}}

\newcommand{\LQ}{``}
\newcommand{\beqry}{\begin{eqnarray}}
\newcommand{\eeqry}{\end{eqnarray}}
\newcommand{\bqn}{\begin{equation}}
\newcommand{\eqn}{\end{equation}}
\usepackage{color}

\begin{document}

\title{From the Standard Model towards a Theory of Matter: Quarks}
\author{T.\ Goldman}\email{tjgoldman@post.harvard.edu}
\author{G.\ J.\ Stephenson,\  Jr.}\email{gjs@stephensonandassociates.com}
\affiliation{Dept. of Physics and Astronomy, University of New Mexico, 
	Albuquerque, NM 87501 \\ 
	{\rm and} \\
	Theoretical Division, MS-B283, 
	Los Alamos National Laboratory, Los Alamos, NM 87545}

\begin{flushright}
\today\\
{LA-UR-19-22186}\\
{arXiv:2019.mm.nnnnn}\\
\end{flushright}

\begin{abstract} 

We follow the example of Cabibbo by revising the Standard Model (SM) to present 
a universal mass structure for fermions. A universal Higgs coupling for each species 
of fundamental fermions moves the SM towards a Theory of Matter, albeit without 
correctly describing the observed mass spectrum. It exposes a need for a complete 
Theory of Matter to include components from physics beyond the Standard Model 
(BSM). Describing the effect of these components phenomenologically provides 
a means to infer the nature of some of the BSM physics required. Our results also 
provide constraints on some BSM matrix elements. Here we apply this concept to 
quarks; the application to leptons will appear in a separate paper.  An immediate 
benefit for theory is the reduction of the largest fine structure constant for the Higgs 
coupling to fermions by an order of magnitude, which improves the perturbative 
appearance of the weak interactions. The small mixing of the third generation 
of each fermion in the fermion families to the others is attributed to the small BSM 
perturbations that produce the small mass ratio of the lighter generations to the 
most massive one.

\end{abstract}

\maketitle			


\section{Introduction}

The Standard Model (SM) consists of two parts: 1) a {\it theory} of interactions, namely the strong 
interactions (QCD),  electroweak theory and Higgs field theory with its vacuum expectation 
value ($vev$), for the origin of the masses of the weak bosons, and 2) a {\it model} of fermions, 
namely a compendium of fermion degrees of freedom that are subject to these interactions 
with a relatively large number of {\it ad hoc} parameters to describe the range of observed 
masses (all still arising from the Higgs $vev$) and the mismatches between the fermion mass 
and weak interaction eigenstates (weak doublets).  This latter part was implicit in the title 
of Weinberg's seminal 1967 paper~\cite{SW}.

We focus on the implication in the SM that there are quantum numbers beyond weak isospin 
and charge  -- additional flavors -- which distinguish among the three generations of families of
fermions. However, none are known despite decades of effort to discern them, and the additional 
flavor quantity that presently defines the distinctions is, in fact, simply mass, for the charged 
fermions. This is made abundantly clear by the fact that this kind of flavor for neutrinos is not 
defined by neutrino mass but rather by the mass of the charged lepton to which the neutrino 
couples. All of this is further emphasized by the misalignment, between the weak interaction 
current isospin doublets and the mass eigenstate pairs, which is described by the 
Cabibbo-Kobayashi-Maskawa (CKM) matrix~\cite{Cab, KM} for quarks and the corresponding 
Pontecorvo-Maki-Nakagawa-Sakakta (PMNS) matrix~\cite{PMNS} for leptons.  

Nonetheless, there certainly must be quantum numbers that distinguish fermions of the same 
charge since, for example, the indefinite mass states constructed by inverting the action of the 
CKM on the massive states, are fermions with left-chiral parts that appear in precisely aligned 
pairs under the weak interaction and the coupling to the Higgs field. These states may properly 
be termed \LQ current" fermions and must have some distinguishing feature so that both the charged 
vector bosons of the SM and the Higgs scalar recognize which pairs couple with unit strength.

 \subsection{Historical Considerations}

Of course, the SM has been exceptionally successful with only a few recent hints of possible 
problems.~\cite{SMprobs} However, it remains a combination of model and theory. We apply 
two lessons from the history of physics in an attempt to resolve this issue.  

One, from particle physics, is the value of insisting upon universality, most famously invoked by 
Cabibbo~\cite{Cab} to solve the problem of the decay rates in kaon decays that are markedly 
different compared to the expectations from Fermi theory as applied in nuclear beta decay and 
pion decay. This ultimately led to Glashow, Illiopoulos and Maiani~\cite{GIM} predicting the 
existence of the charm quark and influenced Kobayashi and Maskawa~\cite{KM} as they 
improved our understanding of CP-violation and predicted the existence of another pair of 
quarks which have also since been identified. 

The other, from nuclear physics, is that a good starting model may have \LQ too much" physics 
in it. As remarked by Weisskopf~\cite{VW}, several early nuclear models suffered from deviating 
more strongly from data when what appeared to be additional correctional elements of physics 
were added. This could be interpreted as being due to some of that physics already being included 
in the initial model. 

Applying both of these lessons produces a theory of luminous (known) matter that does {\em not} 
agree with data as well as the SM, but points the way to contributions from new (BSM) physics 
which can be interpreted as being due to interactions with unknown, non-luminous, dark matter 
which is now known to exist.~\cite{DM}

 \subsection{Application Elements}

As it stands, the SM remains an incomplete theory with model components.  As observed by 
Kaus and Meshkov~\cite{othrKM} almost 30 years ago, one way to advance to a theory is to 
recognize that, although the Higgs field precisely responds to which pairs of  left-chiral, weak 
doublet Weyl spinors are related by the weak interaction, in parallel to the transitions due to 
the charged W-bosons, it has no known means to determine which of these combine with the 
(independent) particular right-chiral, Weyl  spinors that appear in weak singlets. This feature 
of the resulting theory was called \LQ democracy" by Jarlskog~\cite{Jarl}. 

Aspects of this observation were made earlier by a number of authors, notably Fritzsch and 
Planckl.~\cite{FP} However, the corrective efforts involved either assuming a substructure to 
the quarks and leptons or embedding them in larger (\LQ grand unifying") gauge theories to 
acquire quantum numbers to distinguish distinct fermions of the same charge.~\cite{PDGU}

The apparent problem of the \LQ democratic" theory is that it provides mass via the Higgs $vev$ 
to only one mass eigenstate of each triplet of Dirac bi-spinors, of each (electric) charge, that are 
formed from appropriate pairs of the Weyl spinors (as will be displayed below). In the past, this 
was viewed as an impediment to proceeding along this path.  Nonetheless, it has an appeal in 
restoring a kind of universality much as Cabibbo restored the universality of the weak interaction 
strength by inferring that the weak interaction eigenstates consisted of combinations of strong 
interactions mass eigenstates that we now describe in terms of quarks.

More recently, the existence of dark matter (DM) has been amply confirmed~\cite{DM}
and the question of its possible very weak interaction with luminous (known) matter has arisen. 
We note here that this affords an opportunity to reconsider the Kaus-Meshkov~\cite{othrKM}
approach to a theory by considering the possibility that there are perturbatively small corrections 
to the Higgs mass matrices due to very weak interactions of luminous matter with DM. Following 
this theory route allows for the unambiguous identification of some of the BSM physics needed 
if this formulation is correct. We examine this in detail and show that it can indeed provide an 
accounting for all of the observed masses and (weak) mixings using only perturbatively small 
corrections (less than several percent) which should be calculable in any theoretical extension 
that includes DM. Our results provide constraints on those extensions. 

Consistent with the retrenchment from the SM to Higgs Universality plus phenomenological 
additions, the small corrections from the additional physics necessarily involve DM degrees 
of freedom and so illuminate (a part of) that sector. Without assuming detailed knowledge of 
the new physics, but with the assumption of the validity of the see-saw mechanism for neutrino 
masses~\cite{seesaw}, it is also possible to predict, consistent with current experimental 
observations, the approximate range of the ratios of the masses of three sterile neutrinos. 
These predictions constitute a test of this approach as long as the see-saw mechanism is 
also valid. Without it, a different approach would be required. We defer an analysis of 
leptons to a separate paper. 

\section{Weyl-Spinor/Chiral Basis for Democracy}

As demonstrated by the attempts at Grand Unification~\cite{Georgi-Glashow,SO10}, 
mass terms in the SM are well-described as arising from elements of the Lagrangian 
in which a Weyl spinor from a weak interaction (SU(2)) doublet is coupled via the Higgs 
boson to a Weyl spinor that is a weak interaction singlet. The mass term develops when 
the electric charge neutral component of the Higgs boson acquires a $vev$. If the two 
spinors involved are then phase locked together to form a Dirac bispinor, the terms may 
be rewritten into the conventional form of a Dirac mass term. This detail is usually 
obscured by writing the interaction directly in terms of left- and right-chiral projections 
of Dirac bispinors. To be consistent with the symmetries of the theory, this can only 
be done for pairs that have matching combinations of weak isospin and U(1) weak 
hypercharge quantum numbers which produce a conserved electric charge. 

However, within the SM, there are no other quantum numbers to determine which compatible 
pairs of singlet and doublet components should be matched. Conventionally, in the Dirac form, 
one simply adjusts the Higgs coupling so that the product with its $vev$ equals one of the 
experimentally determined values and assigns that to a particular pair in that form. The very 
large value required for the dimensionless coupling for the top quark raises a concern regarding 
the validity of calculations using perturbation theory~\cite{who1}. Furthermore, the mismatch 
between mass and current states that presents itself in the CKM matrix~\cite{Cab,KM}, including 
the existence of CP-violation, is completely {\it ad hoc} (phenomenological). 

To resolve this conundrum, we recall that, by means of his mixing angle, Cabibbo~\cite{Cab} 
resolved the difference of the weak interaction strength for strange hadrons from that for 
non-strange hadrons (including nuclei) as contrasted with the universality of the weak interaction 
for electrons and muons. The requirement of a universal weak interaction for hadrons, {\it i.e.}, 
\LQ Cabibbo Universality", opened the door to both the prediction the charm quark~\cite{GIM} 
and the development of the SM as a whole. Here we propose that it is beneficial to extend that 
universality to include the interactions of the Higgs boson with all of the fundamental fermions 
of each charge type, {\it i.e.}, Higgs Universality. 

The mass spectra themselves invite consideration of such an approach: For all three triples of 
electrically charged fermions, two mass values are considerably smaller than the largest mass, 
which is reminiscent of the \LQ pairing gap" spectrum~\cite{pairing} if the smaller values may be 
approximated as negligible. This spectrum arises from a mass matrix in which all of the entries 
are identical (Higgs Universality), which appears natural in the absence of quantum numbers that 
distinguish which pairs of the Weyl spinors (one from a weak doublet and one from a weak singlet) 
are specifically to be related. 

This observation was essentially first made as much as four decades ago~\cite{othrKM,democ} 
and the corresponding mass matrix was termed \LQ democratic"~\cite{Jarl}. At the time, the 
complete set of quark and charged lepton masses was not known. Now that it is, one can apply 
the inverse of the unitary transformation (which in its conventional form is termed \LQ tri-bi-maximal" 
or TBM) that diagonalizes the democratic matrix and determine the magnitude of the deviations 
from democracy. When scaled by the largest mass value, they are indeed found to be equal to 
each other to within a few percent or less. 

This result supports a conjecture that the deviations from equality are due to perturbative 
corrections from physics beyond the SM which is conventionally termed BSM physics. By 
appropriately scaling and parametrizing these terms, they may be examined for \LQ 
naturalness" and related to the two smaller masses in each triple of Dirac fermions with 
a common electric charge. Loop corrections involving BSM degrees of freedom are a 
plausible origin for such perturbations. A side benefit is the reduction of the largest 
dimensionless coupling between the quarks (of each charge) and the Higgs by a factor 
of three, corresponding to a reduction by an order of magnitude of the largest Higgs 
\LQ fine structure" constant. This reduction improves the perturbative appearance of the 
weak interactions.~\cite{who1}

For the quarks, one may immediately go further as the unitary matrices (which then differ from 
being exactly TBM) that diagonalize the mass matrices also describe the misalignment between 
the mass eigenstates and the weak current eigenstates. The product of the adjoint of that unitary 
transformation for the (electric charge +2/3) up quarks with the one for the (electric charge -1/3) 
down quarks forms the CKM matrix.~\cite{Cab,KM} The CKM now describes the deviations from 
weak interaction universality in the charge-raising quark weak current as also being due to BSM 
physics. In particular, the smallness of the mixing between each of the family members in the 
(so-called) third \LQ generation" 
and those in the other two, due to the large difference in mass, follows immediately. Matching the 
experimental CKM values provides additional constraints on the parameters describing the 
BSM physics and further identifies BSM physics as the source of CP-violation. The unitary 
transformation matrices of both the up quarks, and the down quarks include a TBM  factor, 
which cancels out in the product and leaves the CKM sensitive to the (presumed) BSM corrections. 


In the following sections, we display the calculations that we have described here. 

\section{Quark masses}

Since there is no advantage to using the Weyl spinor formulation for the quarks, we maintain
the Dirac bispinor representation here. 

The conventional TBM matrix
\bqn
TBM  =     \left[ \begin{array}{ccc} 
\frac{1}{\sqrt{6}}   & - \frac{1}{\sqrt{2}}  & \frac{1}{\sqrt{3}}  \\
\frac{1}{\sqrt{6}}   &  \frac{1}{\sqrt{2}}  & \frac{1}{\sqrt{3}} \\
-\frac{2}{\sqrt{6}}  &  0 & \frac{1}{\sqrt{3}} \\
\end{array}
\right]					\label{eq:TBM}
\eqn
diagonalizes the (so-called) \LQ democratic" matrix 
\begin{equation}
M_{dem} =   \frac{1}{3} \times
 \left[ \begin{array}{ccc} 
 1 &  1  &  1  \\
1 &  1  &  1  \\
 1 &  1  &  1 \\
 \end{array}
\right]
\end{equation}
to 
\beqry
M_m & = & TBM^{\dagger} \times M_{dem} \times TBM  \nonumber \\
& = &  \left[ \begin{array}{ccc} 
0 &  0  & 0  \\
0  &  0 &  0  \\
0  & 0 &  1\\
 \end{array}
\right]				\label{eq:TBMtrans}
\eeqry
where we have chosen the overall scale so the nonzero eigenvalue is unity. 

\subsection{Accuracy of a Higgs Universal Initial Mass Matrix }

The accuracy of a Higgs Universality conjecture for quarks can be tested by inverting 
the TBM transformation on the known quark masses (taken from the Particle Data Group 
(PDG)~\cite{PDG} and) placed into diagonal mass matrices for the up quarks and down 
quarks, respectively, {\it viz.},
\bqn
m_u   =  \left[
\begin{array}{ccc}
2.3 & 0 & 0 \\
0 &  1275 &0 \\
0 &  0 & 173500 
\end{array} \right]  \\ 
\eqn
and 
\bqn
 m_d  =  \left[
\begin{array}{ccc}
3.8 & 0 & 0 \\
0 &  95 &0 \\
0 &  0 & 4150 
\end{array} \right]  
\eqn
where all values are expressed in MeV/$c^2$. We will ignore the significant 
uncertainties and variation with scale of these masses\cite{scale} as the ratios 
vary less dramatically, and the values of even the ratios are not known to very 
high accuracy. 

Transforming these inversely using the $TBM$ matrix given in 
Eq.(\ref{eq:TBM}), we see that the resulting mass matrices are indeed 
almost exactly as expected if Higgs Universality is correct:
\beqry
m_{u-TBM} & = & TBM \times m_{u} \times TBM^{\dagger} \nonumber \\
  & = & (173500) \times  \left[
\begin{array}{ccc}
0.33701 & 0.32966 & 0.33333 \\
0.32966 &  0.33701 & 0.33333 \\
0.33333 &  0.33333 & 0.33334
\end{array} \right]   \label{demoup}
\eeqry
and similarly 
\bqn m_{d-TBM}  =   (4150) \times  \left[
\begin{array}{ccc}
0.34493 & 0.32204 & 0.33303 \\
0.32204 &  0.34493 & 0.33303 \\
0.33303 &  0.33303 & 0.33394
\end{array} \right]  				\label{demodn}
\eqn
where we have scaled out the overall factor of the largest mass in each case. 
Although the true accuracy is, of course, far less, we keep the extra digits 
to display which matrix elements are not identical after the (inverse) TBM 
transformation and so convey the patterns that will survive even substantial 
(within experimental uncertainties) changes in the ratios of the diagonal values. 

(We note in passing that equivalent results for  two pairs of $u$- and $d$-quarks 
were presented by Fritzsch and Planckl.~\cite{FP2})

This demonstrates that only perturbatively small BSM corrections to a universal 
starting point (Higgs Universality) are needed. (We have ignored $CP$-violation 
considerations here, but will return to them below.) The deviations from universality 
are exceptionally small, $\le 1$\% in the up quark sector and $< 4$\% in the down 
quark sector (for positive and negative deviations from an average).  It is clear from 
this that something close in structure to $M_{dem}$ (times an overall mass scale, 
$m$) is a reasonable ansatz to consider for an initial mass matrix. (A similar result 
holds for the charged leptons.)

This result confirms that the wide range of quark masses is well described by an 
almost \LQ democratic" mass matrix for each charge set of quarks, leaving only the 
overall scale difference between up quarks and down quarks (and also leptons) 
to be understood. We do not address that difference here.

\subsection{Mass Matrix with BSM Corrections} \label{sub:MBSM}

In the {\em current}  quark basis consistently defined by the Higgs and weak vector 
boson couplings (which is perhaps clearer in the Weyl spinor formulation), the 
universal Higgs plus BSM-corrected mass matrix for each set of 3 quarks of a 
given electric charge has the form 
\bqn
{\mathcal M_{HUB}} = m\times[ {\mathcal M_{HU}}+ {\mathcal M_{BSM}} ]
\eqn
where ${\mathcal M_{HU}} = M_{dem}$ and 
\bqn
{\mathcal M_{BSM}} =   \epsilon \left[ \begin{array}{ccc} 
  \sqrt{\frac{2}{3}}y_0+y_3+\frac{1}{\sqrt{3}}y_8  &  y_1-I{y_2} & y_4-I{y_5} \\
y_1+I{y_2} & \sqrt{\frac{2}{3}}y_0-y_3+\frac{1}{\sqrt{3}}y_8 & y_6-I{y_7} \\
y_4+I{y_5}  & y_6+I{y_7} & \sqrt{\frac{2}{3}}y_0-\frac{2}{\sqrt{3}}y_8 \\
 \end{array} \right]  						\label{eq:bsm}
\eqn

Here, $m$ is an overall scale which is approximately one-third of the mass of the most 
massive of each triple of quarks of a given non-zero electric charge. The BSM correction 
mass matrix, ${\mathcal M_{BSM}}$, accommodates the most general set of deviations 
possible for a Hermitean $3\times 3$ matrix from the democratic mass matrix produced 
by universal Higgs coupling in each quark charge sector. The coefficients are chosen to 
match the normalization of the standard Gell-Mann $SU(3)$ ($U(3)$) basis matrices. 

The BSM corrections are all taken here to be proportional to the small quantity,  $\epsilon$, 
defined by the diagonal matrix of known mass eigenvalues, (again, with the overall scale 
factored out)
\begin{equation}
M_{diag} = {\tilde m} \times  
 \left[ \begin{array}{ccc} 
 \epsilon \delta &  0  & 0  \\
0  &  \epsilon &  0  \\
0  & 0 &  1\\
 \end{array}
\right]	 \label{eq:diag}
\end{equation}
where ${\tilde m}$ may differ from $m$ by a term of ${\mathcal O}(\epsilon)$ which is irrelevant here.


\subsection{Mass Ratio Parameter Values}

As is apparent from Eq.(\ref{eq:diag}), $\delta$ is the ratio of the mass of the lightest mass eigenstate 
of the three quarks (with the same electric charge) to the mass of the intermediate mass quark, and 
$\epsilon$ is the ratio of that quark mass eigenstate to the most massive of the three. In particular, for 
the quark mass values referred to above, 
\beqry
\epsilon_u =   7.35 \times 10^{-3},   \,\,   & & \,\,   \delta_u =   1.8 \times 10^{-3}   \nonumber \\
\epsilon_d =    2.29 \times 10^{-2},   \,\,  & &  \,\,    \delta_d =   4.0 \times 10^{-2}
\eeqry 
Even the largest of these values easily qualifies as a small expansion parameter. We will see below 
that the $\delta$s do not significantly influence our results, so the largest perturbation is provided by 
$\epsilon_d$. 

\subsection{Diagonalization by Unitary Transformation}

One may solve for the eigenvectors and eigenvalues of ${\mathcal  M_{HUB}}$ as functions of the $y_j$. 
(We carried out that approach in an earlier version of this analysis, see Ref.(\cite{usV3}).) However, we 
can directly infer from the required result, Eq.(\ref {eq:diag}), that to leading order in $\epsilon$, the form 
of the unitary matrix $X_{tot}$ that diagonalizes ${\mathcal M_{HUB}}$ via $X_{tot}^{\dagger} \times {\mathcal 
M_{HUB}} \times X_{tot}$ may be structured as  
\beqry
X_{tot} & = & TBM  \times  X_{3\rightarrow2}  \times  R2_{\omega}   \\
& = & TBM  \times  
 \left[ \begin{array}{ccc} 
 1 &  0  & \epsilon x_a e^{\imath \zeta}  \\
0  & 1 & - \epsilon x_b  \\
-\epsilon x_a e^{-\imath \zeta}  & \epsilon x_b &  1\\
 \end{array}
\right]  \times  
\left[ \begin{array}{ccc} 
 {\rm cos}(\omega) &   {\rm sin}(\omega)  & 0  \\
-{\rm sin}(\omega)    & {\rm cos}(\omega) &0  \\
0 & 0 &  1\\
 \end{array}
\right] 	 \label{eq:xtot}
\eeqry
since the TBM factor will diagonalize all of the $\mathcal{O}(1)$ contributions and, for the right values 
of $x_a$ , $x_b$, and $\zeta$, the second factor ($X_{3\rightarrow2}$) will block diagonalize the 
$\mathcal{O}(\epsilon)$ terms to a $2X2$ matrix, which can itself be diagonalized by a simple $2X2$ 
rotation through an angle $\omega$ ($R2_{\omega}$). 

This is a sufficient approximation as the corrections to unitarity in $X_{3\rightarrow2}$ are 
$\mathcal{O}(\epsilon^{2})$ or higher everywhere and $\mathcal{O}(\delta \epsilon^2)$ in 
the (1,2) and (2,1) entries. Thus they do not affect the rotation in the 1-2 plane at an order 
of significance for our calculations here. 
The angle $\omega$ need not 
be small, however. For instance, if the block diagonalization produces a matrix that also has 
almost identical entries, then $\omega \sim \frac{\pi}{4}$ would be required to produce the 
eigenvalues of $\epsilon$ and $\epsilon \delta$. 

We recognize that this approximate block diagonalization approach does not follow the normal Euler
 method, but rather effectively makes a rotation of the 3-axis about a particular axis in the 1-2 plane to 
 a new 3-axis slightly [$\mathcal{O}(\epsilon)$] tilted over the 1-2 plane, followed by a rotation about the 
 new 3-axis produced. We have checked that, as expected, this produces the same results at each 
 level of approximation as a sequence of small Euler rotations, first about the 1-axis, then the 2-axis 
 and finally by a not necessarily small rotation about the 3-axis. Both methods can be taken to high 
 orders of $\epsilon$ leaving only arbitrarily small deviations from exact diagonalization of the mass 
 matrix and from unitarity of the resulting $X_{tot}$.

By applying the inverse of this transformation to $M_{diag}$ and expanding to $O(\epsilon)$, 
we can find ${\mathcal M_{HUB}}$ in terms of only the four unknowns, $x_a$ , $x_b$, $\zeta$ 
and $\omega$. By projecting both the ${\mathcal O}(\epsilon)$ parts of this resulting matrix 
and of ${\mathcal M_{BSM}}$ using the nine standardly normalized Gell-Mann matrices, we 
determine that the nine $y_j$ parameters are not completely independent and can be defined 
in terms of only these four parameters. (In fact, $\mathcal{O}(\epsilon^2)$ corrections to the 0 
and 1 entries suffice to promote unitarity uniformly up to $\mathcal{O}(\epsilon^3)$ without 
additional parameters.) We find
\beqry
y_0 & = & \frac{1}{\sqrt{6}} (1 + \delta) \nonumber \\
y_3 & = & \frac{1}{2\sqrt{3}} [2\sqrt{2}x_b - (1 - \delta){\rm sin}(2\omega)] \nonumber \\
y_8 & = & \frac{1}{2\sqrt{3}} [2\sqrt{2}x_a {\rm cos}(\zeta) + (1 - \delta){\rm cos}(2\omega)] \nonumber \\
y_1 & = & \frac{1}{6} [ 2\sqrt{2}x_a {\rm cos}(\zeta) - 2{\rm cos}(2\omega)(1 - \delta) - (1 + \delta) ] \nonumber \\
y_4 & = & \frac{1}{6} [ -\sqrt{2}x_a {\rm cos}(\zeta) +\sqrt{6}x_b + 
	({\rm cos}(2\omega)+\sqrt{3}{\rm sin}(2\omega))(1 - \delta) - (1 + \delta) ] \nonumber \\
y_6 & = & \frac{1}{6} [ -\sqrt{2}x_a {\rm cos}(\zeta) - \sqrt{6}x_b + 
	({\rm cos}(2\omega)-\sqrt{3}{\rm sin}(2\omega))(1 - \delta) - (1 + \delta) ] \nonumber   \\
y_2 & = & 0 \; , \;  y5 = y7 = -\frac{1}{\sqrt{2}} x_a {\rm sin}(\zeta) 	\label{eq:constraint}
\eeqry

In principle, these nine $y_j$ are independent quantities for each type of charged fermion, in 
addition to the two already experimentally (approximately) known quantities, $\epsilon$ and 
$\delta$. 

Eqs.(\ref{eq:constraint}) constrain the freedom of proposed BSM models and demonstrate 
that only four $y_j$ (for each type of charged fermion) can be independent (as well as $\epsilon$ 
and $\delta$) in any proposed BSM model, as any deviations from the relations presented here must 
be higher order small, or the putative BSM model must be incorrect. These relations do have higher 
order corrections, but calculation of them is not warranted at this time, given that many of the quark 
masses are poorly known at present. Furthermore, as we will see below, with the exception of $y_0$ 
which is uniquely determined in terms of $\delta$, only differences between the parameters for the 
pairs of quark types can be determined from the CKM mixing matrix. 

We note that this construction sets $y_2$ to zero. This is essential, but is most easily understood by 
proceeding in the opposite direction, by starting with the $y_j$ and constructing $X_{tot}$ from them. 
It is then clear that if $y_2 \ne 0$, there will be large $CP$-violation in the light quark sector as the 
factor of $\epsilon$ factors out as an overall factor in the $2 \times 2$ rotation, leaving $\mathcal{O}(1)$ 
$CP$-violation terms (as we showed in Ref.(\cite{usV3}) and repeat here later). We postpone further 
discussion of $CP$-violation until we have developed the explicit CKM matrix. 

We also note in advance that the Cartan subalgebra terms alone are insufficient to fit the experimental 
results for the CKM for the small values of $\delta$ seen above.  A large value of $\delta$ would obviate 
the entire approach. We similarly postpone further discussion of this until after we develop the explicit 
CKM matrix. 

\section{Fitting to the CKM matrix}  \label{fitz}

To compute the CKM matrix, we need the result in Eq.(\ref{eq:xtot}) evaluated for both the up quarks, 
and the down quarks. The separate matrices for these are conventionally labelled $U$ and $V$ 
respectively~\cite{PDG}, so that
\bqn
CKM = U \times V^{\dagger}.  \label{CKMdefn}
\eqn

However, the PDG description is one in which these matrices transform from mass eigenstates to 
current eigenstates, but our derivation above is for the transformation of current eigenstates to mass 
eigenstates. Hence, the Hermitian conjugates are interchanged and the $V^{\dagger}$ of the PDG is 
our $X_{tot}$ for the down quarks and similarly, their $U$ is the Hermitian conjugate, $X_{tot}^{\dagger}$,  
of our $X_{tot}$ for the up quarks. 

On combining the results for up quarks and down quarks to produce the equivalent of the CKM 
matrix, 
\bqn
CKM_{0} =  R2_{\omega u}^{\dagger} \; X_{3\rightarrow2 u}^{\dagger} \; TBM^{\dagger} \; TBM \; 
                                    X_{3\rightarrow2 d} \; R2_{\omega d}  \label{eq:CKMform0}
\eqn
we see that the $TBM$ factor cancels out in the product. (We will see for leptons that a more complex 
development is both required and available in the see-saw mechanism, to match the mixing matrix for 
neutrinos.)  Hence, it is sufficient to calculate 
\beqry
CKM_{1} & = &  R2_{\omega u}^{\dagger} \; X_{3\rightarrow2 u}^{\dagger} 
                        X_{3\rightarrow2 d} \; R2_{\omega d}  \nonumber \\
   & = &    \left[ \begin{array}{ccc} 
 {\rm cos}(\Theta_C)   &   {\rm sin}(\Theta_C)  &CKM_1(1,3)  \\
-{\rm sin}(\Theta_C)   &   {\rm cos}(\Theta_C)  & CKM_1(2,3)  \\
CKM_1(3,1) & CKM_1(3,2) &    1 \\
\end{array}   \right]     \label{eq:CKMform1}      
\eeqry                                           
where we have taken advantage of simple trigonometric relations to rewrite the $2\times2$ block 
in terms of the angle difference, $\omega_d - \omega_u  =  \Theta_{C}$, {\it i.e.}, (approximately) 
the Cabibbo angle. 

The other elements are
\beqry
CKM_1(1,3)   & = &   (\epsilon_{d} x_{ad}e^{\imath\zeta_d}-\epsilon_{u} x_{au}e^{\imath\zeta_u}){\rm cos}(\omega_u) 
+ (\epsilon_{d} x_{bd}-\epsilon_{u} x_{bu}){\rm sin}(\omega_u)		   \\
CKM_1(2,3)   & = &   - (\epsilon_{d} x_{bd} - \epsilon_{u} x_{bu}){\rm cos}(\omega_u) 
+ (\epsilon_{d} x_{ad}e^{\imath\zeta_d}-\epsilon_{u} x_{au}e^{\imath\zeta_u}){\rm sin}(\omega_u)  \\
CKM_1(3,1)   & = &   - (\epsilon_{d} x_{ad}e^{-\imath\zeta_d} - \epsilon_{u} x_{au}e^{-\imath\zeta_u}){\rm cos}(\omega_d)
 -(\epsilon_{d} x_{bd}-\epsilon_{u} x_{bu}){\rm sin}(\omega_d) \\
CKM_1(3,2)   & = &   (\epsilon_{d} x_{bd} - \epsilon_{u} x_{bu}){\rm cos}(\omega_d)	 
- (\epsilon_{d} x_{ad}e^{-\imath\zeta_d}-\epsilon_{u} x_{au}e^{-\imath\zeta_u}){\rm sin}(\omega_d) 
\eeqry
which demonstrates that the $CKM$ mixing depends solely on the difference between the diagonalizations 
of the up quarks and the down quarks, as it must. By defining
\beqry
 X  & = &   -(\epsilon_{d} x_{bd} -\epsilon_{u} x_{bu} ) 				 \label{eq:CKMX}  \\
 Y  & = &    \epsilon_{d} x_{ad}{\rm cos(\zeta_d)} - \epsilon_{u} x_{au}{\rm cos(\zeta_u)} \label{eq:CKMY} \\
 Z  & = &    \epsilon_{d} x_{ad}{\rm sin(\zeta_d)} - \epsilon_{u} x_{au}{\rm sin(\zeta_u)} \label{eq:CKMZ}
\eeqry
these elements may be simplified somewhat to the form 
\beqry
CKM_1(1,3) &=& Y{\rm cos}(\omega_u)-X{\rm sin}(\omega_u)+\imath Z{\rm cos}(\omega_u)  \\
CKM_1(2,3) &=& Y{\rm sin}(\omega_u)+X{\rm cos}(\omega_u)+\imath Z{\rm sin}(\omega_u)  \\
CKM_1(3,1) &=& -Y{\rm cos}(\omega_d)+X{\rm sin}(\omega_d)+\imath Z{\rm cos}(\omega_d)  \\
CKM_1(3,2) &=& -Y{\rm sin}(\omega_d)-X{\rm cos}(\omega_d)+\imath Z{\rm sin}(\omega_d) 
\eeqry  

This result still does not match the form of the CKM matrix of the PDG (see below), as it has a 
nonzero imaginary term in the $(2,3)$ matrix element as well as in the $(1,3)$ matrix element. 
This may be remedied by making use of the same phase freedoms that are used to put the CKM 
matrix of the PDG into its standard form. There are six phases available in the CKM matrix, three %
each from the up quark and down quark sectors. We have implicitly used two of each of the three 
available in each mass matrix to reduce the form of $X_{tot}$ in Eq.(\ref{eq:xtot}) to have only one 
phase each for the up quarks and down quarks. Of the remaining two, one is an overall phase which 
can have no effect. We use the last one by choosing for it the value defined by
\bqn
\Phi  = {\rm arctan}\left(\frac{Z{\rm sin}(\omega_u)}
{X{\rm cos}(\omega_u) +Y{\rm sin}(\omega_u)}\right)
\eqn

Upon multiplying $CKM_1$ from the right by the phase matrix
\bqn
PHS =   \left[ \begin{array}{ccc} 
1  &   0  & 0  \\
0   &   1  &  0  \\
0 & 0 &   e^{-\imath \Phi} \\
\end{array}   \right] 
\eqn
and on the left by its adjoint, we obtain our final form of the mixing matrix 
\bqn
CKM_{BSM} =    \left[ \begin{array}{ccc} 
 {\rm cos}(\Theta_C) & {\rm sin}(\Theta_C) & {\mathcal R}[(1,3)]+ \imath \frac{XZ}{Q}\\
-{\rm sin}(\Theta_C)   &   {\rm cos}(\Theta_C)  & Q  \\
 {\mathcal R}[(3,1)]+\imath \frac{XZ}{Q}{\rm cos}(\Theta_C)  &   {\mathcal R}[(3,2)]+\imath \frac{XZ}{Q}  {\rm sin}(\Theta_C) &    1 \\
\end{array}   \right]     \label{eq:CKMBSM}      
\eqn
where 
\bqn
Q  = \sqrt{(X{\rm cos}(\omega_u)+Y{\rm sin}(\omega_u))^2 +(Z{\rm sin}(\omega_u))^2} \label{eq:Q}
\eqn
is real ($ {\mathcal R}$) and the other real parts are 
\beqry
 {\mathcal R}[(1,3)] &= & \frac{ (2{\rm cos}(\omega_u)^2 -1)XY -{\rm sin}(\omega_u){\rm cos}(\omega_u)(X^2-Y^2-Z^2)}{Q} 
 				\label{eq:RE13}  \\
 {\mathcal R}[(3,1)] &= & \frac{X^2{\rm cos}(\omega_u){\rm sin}(\omega_d) -(Y^2 +Z^2){\rm sin}(\omega_u){\rm cos}(\omega_d)-XY{\rm cos}(\omega_u+\omega_d)}{Q} \\
 {\mathcal R}[(3,2)] &= & - \frac{X^2{\rm cos}(\omega_u){\rm cos}(\omega_d) +(Y^2 +Z^2){\rm sin}(\omega_u){\rm sin}(\omega_d)+XY{\rm sin}(\omega_u+\omega_d)}{Q}
\eeqry
 As always, all elements are only shown to first order in the small quantities. (Recall here 
 that $\omega_d = \omega_u + \Theta_C$, so there is only one free angle variable in these 
 formulas.)

\subsection{PDG evaluation}

The PDG~\cite{PDG} provides only the absolute values of the entries of the $CKM$ matrix. 
It also presents a matrix form that has only real entries in the first row and third column of 
the $CKM$ matrix, except for the $(1,3)$ matrix element. This is achieved by locating the 
one required phase in the matrix that produces rotation about the second axis, where the 
sequence of rotations is first about the third axis (which is almost identical with the Cabibbo 
rotation), next about the second axis, and finally about the first axis, proceeding from right to 
left in the products in the usual way, {\it viz.}
\beqry
CKM_{PDG} & =  &  \left[ \begin{array}{ccc}
1 &  0   & 0  \\ 0   & c23 & s23 \\ 0 & -s23   & c23 \\ \end{array} \right] 
\times 
 \left[ \begin{array}{ccc}
c13 &  0   & s13\, e^{-\imath \chi}  \\  0   &1 & 0 \\ -s13\, e^{\imath \chi} & 0   & c13 \\ \end{array} \right] 
\times 
 \left[ \begin{array}{ccc}
c12 &  s12   & 0  \\ -s12   & c12 & 0 \\ 0 & 0   & 1  \\ \end{array} \right]   \nonumber \\
& =  & \left[ \begin{array}{ccc}
c13\,c12 &  c13\,s12   & s13\, e^{-\imath \chi}  \\ 
-c13\,s12 -c12\,s23\,s13\, e^{\imath \chi}   & c23\,c12 -s12\,s23\,s13\, e^{\imath \chi}  & s23\,c13 \\ 
-c12\, s13\, e^{\imath \chi} -c12\,c23\,s13\, e^{\imath \chi}  & -s23\,c12 -s12\,c23\,s13\, e^{\imath \chi}  & c23\,c13 
\end{array} \right] 
\eeqry
where as usual, $c13 = {\rm cos}(\theta_2)$, etc. and we have changed the PDG phase 
notation from $\delta$ to $\chi$ to avoid confusion with our mass ratio parameter above. 

Our construction above agrees precisely with the PDG structure through first order in 
$\epsilon$ as the sines of all of the angles other than $\Theta_C$ are small (see below), 
as they must be of order $\epsilon$ to be consistent with the diagonalization matrices that 
we have constructed for the quarks. We will see immediately that this requirement is satisfied.  

To proceed, we need to have explicit real and imaginary components for the matrix entries, rather 
than only moduli as reported by the PDG. Therefore we have constructed a version of the 
PDG result where we assume that all three of the mixing angles reside in the first quadrant. 
This is not justified, but demonstrates how the constraints on BSM parameters may be extracted 
were such information available and enables us to demonstrate that an acceptable fit solution 
does exist. 

For completeness, we report here the values of the sines of the angles that we have extracted 
by the procedure just described: 
 \beqry
{\rm sin}(\theta_{13})  & = & 0.00357 \nonumber  \\ 
{\rm sin}(\theta_{23})  & = & 0.04110 \nonumber \\
{\rm sin}(\theta_{12})  & = & 0.22506  \nonumber  \\
{\rm sin}(\chi)  & = & 0.94563  \label{eq:sinvals}
\eeqry
We have taken $\theta_{12}$ to be indistinguishable from $\Theta_C$ at this level of accuracy. 
We have also evaluated the $CP$-violation phase using the value of the Jarlskog~\cite{CPV}
invariant (see below). 

Taking these values and using the PDG~\cite{PDG} parametrization, we obtain 
central values for the real and imaginary parts of these quantities, {\it viz.}: 
 \beqry
CKM13  & = & 0.001161 - 0.003376 \, I  \nonumber  \\ 
CKM23  & = &  0.04110 + 0.0 \, I \nonumber \\
CKM31  & = & 0.008120 - 0.003286 \, I \nonumber  \\
CKM32  & = & -0.04031 - 0.0007591 \, I  \label{eq:CKMelvals}
\eeqry
where the {\it rhs} in each case is the corresponding entry of the matrix $CKM_{PDG}$ 
when, as noted above,  the particular set of signs for the sines is chosen corresponding 
to all three angles  being in the first quadrant. Also, using the entries in the upper left 
$2\times2$ block, (which are real through first order in small quantities as $s13$ and 
$s23$ are both small) we estimate the value of the Cabibbo angle, $\Theta_C$, as
\bqn
 \Theta_C = 0.2270  \label{eq:thetaCval} 
\eqn
{\it i.e.,} approximately $13.0^{o}$. 

Other choices for extracting the full matrix elements could be investigated as well, but this 
demonstrates that at least one solution exists. We have investigated a number of alternatives 
and find that the largest differences, apart from signs, are in the real and imaginary parts of 
$CKM13$ and part of $CKM32$, but the changes are not large, e.g., $\sim 20$\%. However, 
if the phase is placed in an alternate location, for example so that the first row entries are {\it all} 
real, then larger changes are obtained in the real and imaginary parts, although the moduli are 
maintained, of course. 

We uniformly present 4-digit values for consistency, but the changes between the 2012, 2014 and 
2016 PDG reports suggest that in a number of cases the values are not known to better than two digits, 
at most, although some of the uncertainties are a small fraction of a percent. On the basis of these 
larger uncertainties, we conclude that carrying out our analysis to order $\epsilon^2$ is not warranted 
at this time. Finally, we note that our numerical representation of $CKM_{PDG}$ {\bf is unitary} to 
better than one part in $10^{10}$.

\subsection{BSM parameter constraints}

\begin{figure}  [h]  
\includegraphics[width=0.8\textwidth, height=0.8\textwidth, angle=0]{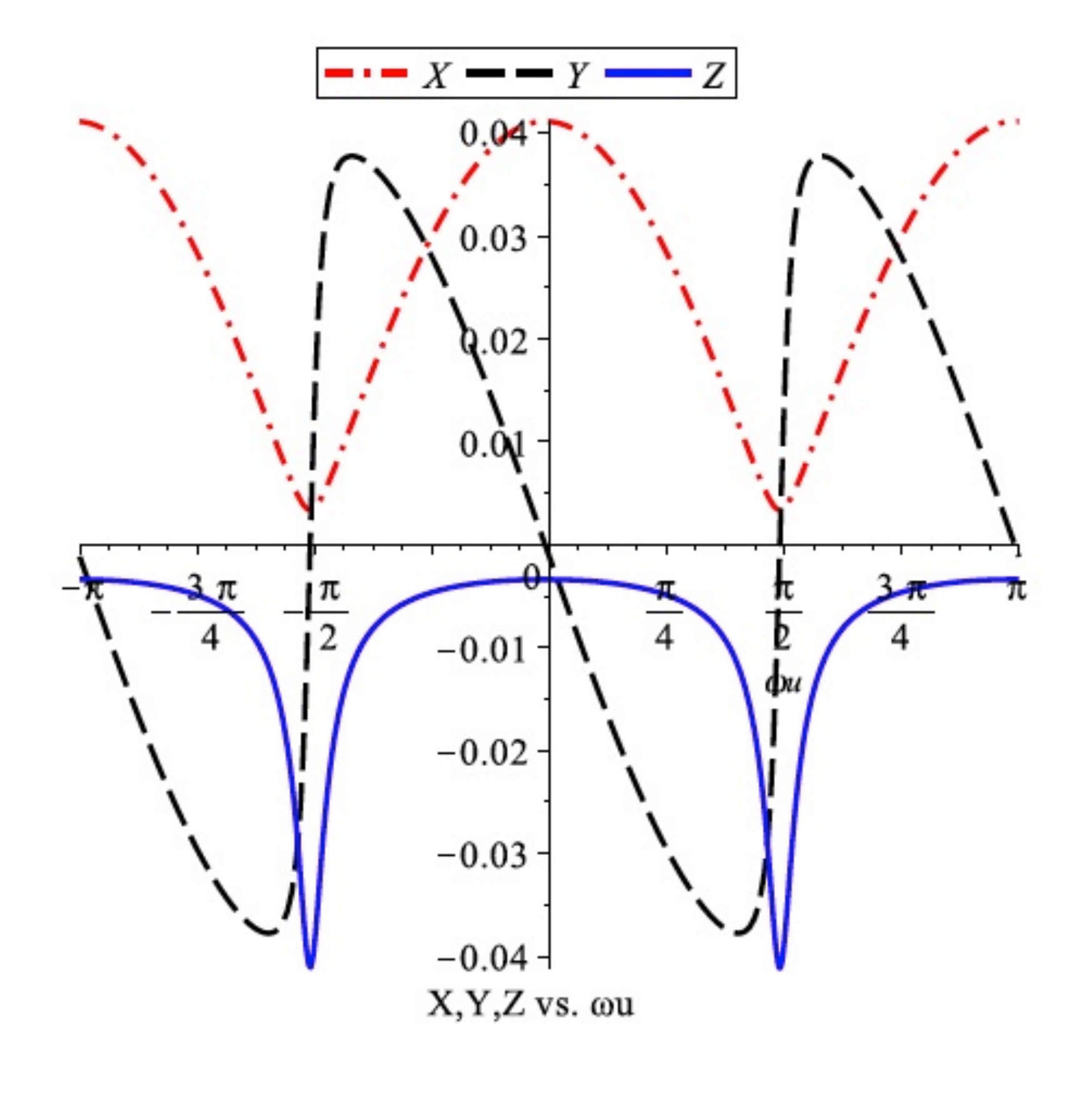}
\caption{Variation of BSM parameter combinations as functions of $\omega_{u}$.}
\label{fig:XYZ}
\end{figure}

We can solve for the values of $X$, $Y$ and $Z$ as functions of $\omega_u$ (using $\omega_d = 
\omega_u + \Theta_C$) by requiring agreement between the values extracted above for the real 
and imaginary parts of $CKM13$ and the real value of $CKM23$ and the functional forms for these 
elements as given in Eqs.(\ref{eq:Q},\ref{eq:RE13}). Using the central values labelled as
\beqry
F  & = & {\mathcal Re}(CKM13) = 0.001161 \\
G & = & {\mathcal Im}(CKM13) = - 0.003376 \\
H & = &   Q =   CKM23  =    0.04110 
\eeqry
we find 
\beqry
X  & = &  \sqrt{(F {\rm sin}(\omega_u))^2 + (G {\rm sin}(\omega_u) - H {\rm cos}(\omega_u))^2}   \label{eq:solnX} \\
Y & = &  \frac{(F^2+G^2-H^2){\rm sin}(\omega_u){\rm cos}(\omega_u)+GH({\rm sin}(\omega_u)^2 - {\rm cos}(\omega_u)^2)}
{\sqrt{(F {\rm sin}(\omega_u))^2 + (G {\rm sin}(\omega_u) - H {\rm cos}(\omega_u))^2}} \\
Z & = & \frac{HG}{\sqrt{(F {\rm sin}(\omega_u))^2 + (G {\rm sin}(\omega_u) - H {\rm cos}(\omega_u))^2}} \label{eq:solns}
\eeqry
These are plotted in Fig.(\ref{fig:XYZ}) for these values of $F$, $G$ and $H$. Note that these quantities 
include factors of $\epsilon$ and so must be of that order for the results to be \LQ natural", {\it i.e.}, 
no parameters much larger than ${\mathcal O}(1)$ are required from the BSM physics to achieve 
the scale of these values. Fig.(\ref{fig:XYZ}) shows that indeed they satisfy this constraint for all values 
of the unconstrained angle.

We have checked that all of the CKM matrix elements are consistent, that the fitted CKM matrix 
is unitary to a high accuracy (better than 1 part in $10^{9}$), and note that the moduli can {\em not} 
all be fitted {\em without} including a contribution from imaginary terms. This last means both that 
Higgs Universality fails without complex contributions from the BSM physics and conversely, that it 
is consistent to view BSM physics as the origin of CP violation, which we discuss next. 

\subsection{CP Violation}

Here, we examine what $CP$-violating implications are introduced by the BSM parameters that we have 
introduced that produce complex amplitudes.  In particular, we examine whether this is sufficient to be the 
only source of $CP$-violation. 

The invariant characterization of $CP$-violation was described by Jarlskog~\cite{CPV}. The Jarlskog 
invariant quantity, which we label ${\mathcal J}$, appears only at order $\epsilon^2$, and is given 
by~\cite{PDG}
\beqry
{\mathcal J}  & = & {\mathcal Im} [ CKM_{i,j} CKM_{k,l} CKM^*_{i,l} CKM^*_{k,j} ]\nonumber \\
& = & {\rm cos}(\theta_{12}) {\rm sin}(\theta_{12}) {\rm cos}(\theta_{23}) {\rm sin}(\theta_{23}) 
{\rm sin}(\theta_{13}) {\rm cos}^2(\theta_{13}) {\rm sin}(\delta) \nonumber \\
& = &  (3.06 \pm 0.21) \times 10^{-5}
\eeqry
up to an overall sign ambiguity, in the standard PDG representation of the CKM matrix. (We 
used this value above to extract the $CP$-violating phase angle in the CKM matrix.) 

At the first order in $\epsilon$ level of approximation, only the combination of matrix elements 
[$i=j=2,k=l=3$] reproduces the correct result for ${\mathcal J}$:
\beqry
{\mathcal J_{2233}} & = &  \pm
{\mathcal Im} [ CKM(2,2) CKM(3,3) CKM^*(2,3) CKM^*(3,2) ]\nonumber \\
 & = &   \pm {\rm cos}(\Theta_C) \times 1 \times  Q \times \frac{XZ}{Q} {\rm sin}(\Theta_C)	\nonumber \\
 & = & \pm  {\rm cos}(\Theta_C)  {\rm sin}(\Theta_C) XZ \\
 & = & \pm  {\rm cos}(\Theta_C)  {\rm sin}(\Theta_C) GH    \label{eq:J2233form}
\eeqry
where we have made use of the entries in Eq.(\ref{eq:CKMBSM}) and the solutions for $X$ and $Z$ 
in Eqs.(\ref{eq:solnX} ,\ref{eq:solns}). We have checked that completing the unitary structure of $X_{tot}$ 
to second order in $\epsilon$ reproduces the correct result from any $i,j,k,l$ combination, but it is, of course, 
more convenient to acquire the result from the first order terms. With the value of ${\mathcal J}$ known, 
this provides a check on one pair of the combined parameters: We find that our fit agrees with the PDG 
value for the modulus of ${\mathcal J}$ to an accuracy of order 1 part in $10^{4}$.

It is straightforward to see from the imaginary parts that the $(3,1)$ and $(3,2)$ elements contain no 
new information beyond that from the $(1,3)$ and $(2,3)$ elements, which is true for the 
real parts also. It is also clear that the imaginary parts are consistent with the form of ${\mathcal J}$ 
as given in Eq.(\ref{eq:J2233form}). 

\section{Discussion}   

The most striking result of the analysis presented here is that the smallness of the 
non-Cabibbo mixing is directly related to the ratio of the middle to largest masses 
of the quarks. Both features are due to the perturbative size of BSM corrections 
to the initial \LQ democratic" starting point. In contrast, the relatively large size of 
the Cabibbo mixing is allowed by the diagonalization process. The size of the 
separate rotations needed to diagonalize the lighter pairs of up and down quarks 
separately are at least convention dependent on the initial choice of axes in their 
two-dimensional subspace. (Of course, this may be constrained to a specific 
orientation in any particular BSM theory.) 

A perhaps surprising result is that the BSM perturbations need not include all 
Cartan sub-algebra components, contrary to common analyses, while conversely, 
non-Cartan sub-algebra BSM perturbations {\em are} required and not only to 
induce $CP$-violation. This can be seen by setting $\zeta$ to zero and then solving 
the equations $y_1=0$, $y_4=0$ and $y_6=0$ for $x_a$, $x_b$, and $\omega$ 
which determines the values of $y_3$ and $y_8$ (where $y_0$ has been fixed 
independently by $\delta$). Combining these in $Q$ (and noting that $Z=0$ for 
$\zeta=0$) shows that, even for the largest value of $\epsilon$ available, $Q < 0.02$ 
which is a factor of two too small compared to the experimental result. Conversely, 
it is possible for the required value of $Q$ to be attained with ${y_3} = 0$ or ${y_8} 
= 0$, and perhaps even both. Even with maximal constraints, only $y_5 = y_7  
\sim 2$ is  required, which is still \LQ natural", (while $y_{4}-y_{6}$ provides only 
a small contribution). 

The mass ratios of the quarks are scale dependent, and one could examine 
the effects of that scale dependence on the CKM matrix and our fit. However, 
even the ratios are generally not that well known and do not vary significantly 
with scale\cite{scale} over the range from 2 GeV, where the lightest quark 
masses are generally defined and determined, to the scale of the $b$-quark, 
nor from there to the weak scale which is also very close to the top quark mass. 
Refinements responding to these issues are certainly warranted, but we do not 
expect them to produce large corrections to the BSM parameter constraints 
determined here. In fact, since the effects considered here are dominated by 
the values of $\epsilon$, only the uncertainties associated with the masses of 
the strange and charmed quarks should be significant, as the $b$- and $t$-quark 
masses are relatively accurately known. Fortunately, the very large uncertainties 
associated with the ratios of the two lightest quarks do not play a significant role 
in establishing the configuration, although they will be important for precision 
analyses. 

We have carried out the straightforward extension of our results to the next higher 
order in $\epsilon$ which might, in principle, be able to further constrain the values 
of the unknown parameters. Unfortunately, utilization requires knowledge of the 
relevant experimental values to order $\epsilon^2$, i.e., to of order a few parts 
in $10^4$, which is an accuracy generally not presently available. More accurate 
measurements could certainly change this conclusion.

\subsection{Current quarks}

The structure we find for the CKM also illuminates our statements about current 
quarks and their relation to the weak currents and Higgs couplings. If what we 
identify as BSM corrections were not present, then the cancellation of TBM factors 
would leave the CKM as an identity after transforming to the (\LQ flavor") mass basis 
as it was in the original \LQ current" basis. This makes it apparent that it does not 
matter what current quark basis is implemented, not withstanding the difference 
between the overall mass scales for up quarks and for down quarks, as long as the 
mass matrices have identical structures. On the one hand, there is nothing in the SM 
itself to require a difference and on the other, the large separation of the individual 
masses of each charge means that it is possible to find a current quark basis where 
the mass matrices are democratic in the absence of BSM corrections. So we may 
conclude that a current quark basis does exist in which the mass matrices are 
democratic. Hence, that basis is available for our starting point as implemented 
here. 

\subsection{BSM contributions}

\begin{figure} 
\includegraphics[width=0.9\textwidth, height=0.6\textwidth, angle=0]{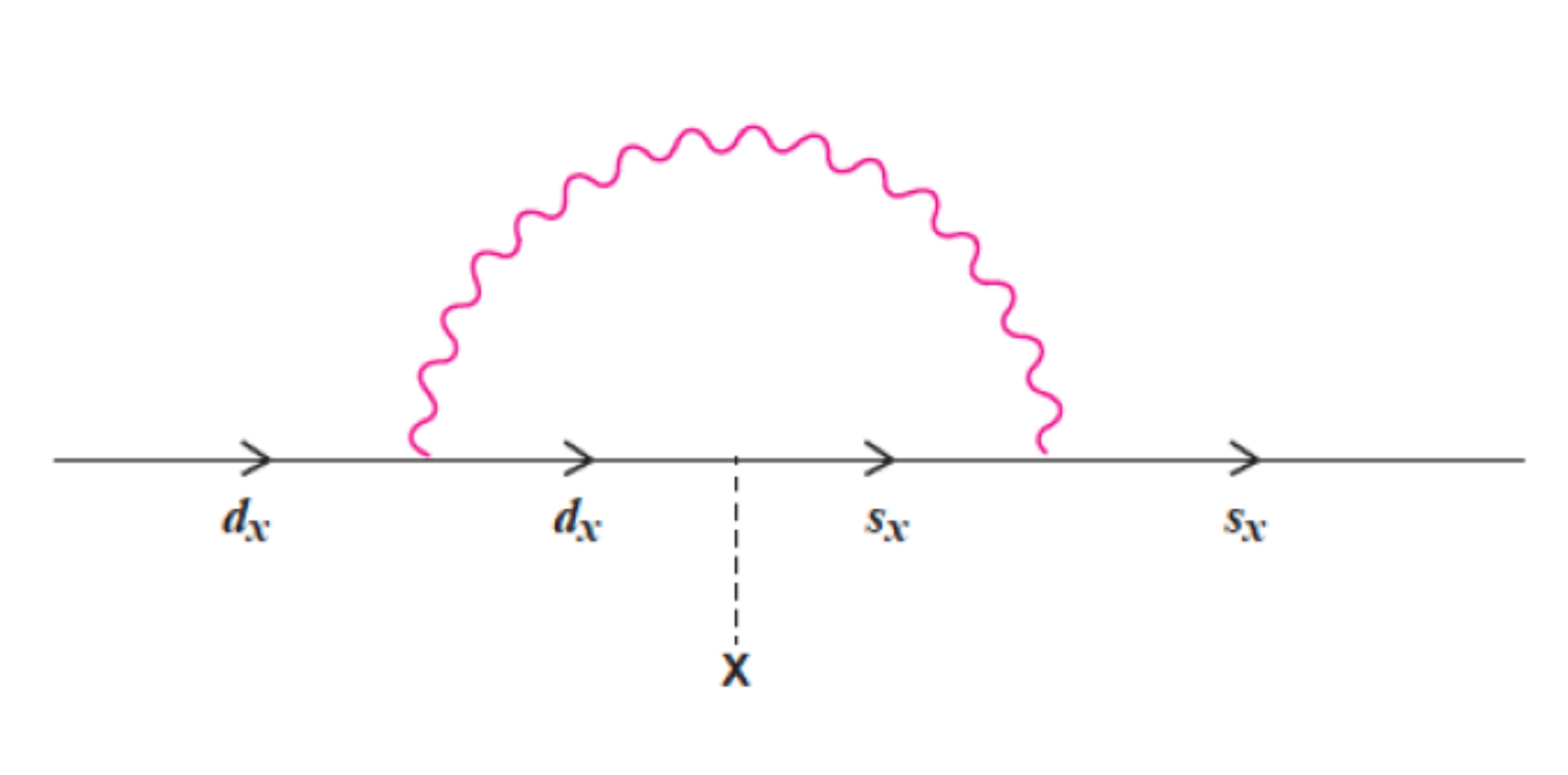}
\caption{A BSM loop correction to the fermion-Higgs-boson vertex that alters 
the charged fermion mass matrix from \LQ democratic" to that shown in Eq.(\ref{eq:bsm}). 
Left-chiral fermions with weak interactions are labelled $d_x$ for \LQ doublet" and 
right-chiral fermions with no weak interactions are labelled $s_x$ for \LQ singlet" 
or \LQ sterile". }
\label{fig:chgdfermions}
\end{figure}

In general terms, Fig.(\ref{fig:chgdfermions}) shows the nature of expected BSM 
corrections that do distinguish the different fermions and could lead to the small 
corrections that we find in our fits. The Lagrangian structure that we have in mind 
uses Weyl spinors for the separate left-chiral ($d_x$, for member of a weak 
interaction doublet) and right-chiral ($s_x$, for a weak interaction singlet) parts 
of the fermion Dirac bispinors, but nonetheless produces Dirac mass terms which 
may be simply represented as above. 

(DM interacting with what we call luminous matter, at least with neutrinos~\cite{intDM}, 
as well as self-interacting DM~\cite{selfintDM}, has already been proposed to solve 
inconsistencies observed if DM had only gravitational interactions, as discussed in   
Ref.\cite{intDM}.)

Interestingly, we expect these loop calculations to be finite as they involve only 
differences within the triples of fermions. This effect was observed in Ref.\cite{GG},  
where a symmetric, overall divergence appears but the differences in mass corrections 
are finite, although that model is for a quite different application of symmetry-breaking 
mass corrections. Also, the calculation there was done only for Cartan sub-algebra 
corrections, but since off-diagonal corrections simply refer to a different Cartan 
sub-algebra, those corrections should also be finite. 

Fig.(\ref{fig:chgdfermions}) is drawn for a BSM gauge vector boson interaction, but the BSM vector 
could in principle also couple the Weyl spinor $d_x$ to a ($CP$-conjugate of the) 
Weyl spinor $s_x$, which simply requires interchanging the labels on either side 
of the Higgs coupling to complete the loop. In that latter configuration of labels 
(without the $CP$-conjugation), the figure could also apply to a BSM scalar boson 
in the loop as well. 

Note that without the intermediation of the Higgs scalar vacuum expectation value,
 both the $d_x$ and the $s_x$ pass through the loop unchanged and are not coupled 
 to mass, so the BSM correction would only affect vertex renormalization. 
 
Finally, we note that the $\epsilon$ parameter is a measure of the coupling of the BSM 
physics to luminous matter, but that in some sense, the $y_j$ represent the intrinsic 
strength of the BSM physics. Since the $CP$-violating $y_2$, $y_5$ and $y_7$ are not 
all zero and $y_5$ and $y_7$ are of $\mathcal{O}(1)$, this suggests that $CP$-violation 
in the BSM physics is not suppressed and invites thinking about the possibility that 
it is \LQ maximal" in some sense, with the usual caveats about how to understand 
that. However, that clearly has significant implications for the development of baryon 
asymmetry in the early Universe as it means that the BSM $CP$-violation is not small, 
as in the SM. This provides alternate routes to large $CP$-violation for the development 
of that asymmetry.

\section{Acknowledgments}
This work was carried out in part under the auspices of the National Nuclear Security
Administration of the U.S. Department of Energy at Los Alamos National Laboratory
under Contract No. DE-AC52-06NA25396. We thank Bill Louis, Geoff Mills (deceased), 
Richard Van de Water, Dharam Ahluwalia, Steve Ellis, Alan Kostele{\' c}ky, Earle Lomon, 
Rouzbeh Allahverdi, Kevin Cahill, Ami Leviatan and Xerxes Tata for useful conversations.

\end{document}